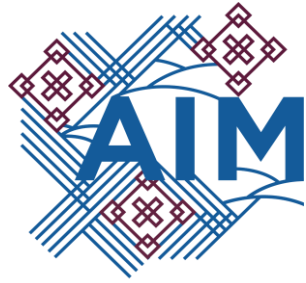

DIJON 2023# Les leaders facilitent-ils l'appropriation des systèmes d'information collaboratifs dans les équipes ?

Laurent Chevalier, CEREGE - Université de Poitiers, France
Thomas Stenger, CEREGE - Université de Poitiers, France
Pierre Laniray, Paris Dauphine, France**Résumé :**

Cette recherche, focalisée sur les petits groupes de travail, propose d'examiner le rôle du leader dans l'appropriation des systèmes d'information collaboratifs. Elle s'appuie sur une recherche-action de trois ans menés lors de la mise en place de la plateforme collaborative Microsoft Teams dans une université. Il s'agit de comprendre comment s'organisent des groupes d'étudiants pour réaliser des projets complexes sur un temps long et avec des contraintes d'éloignements variables. Des différences dans l'appropriation du nouvel outil sont apparues selon que le mode de travail était coopératif ou collaboratif. Pour comprendre ces différentes dynamiques, nous mobilisons les théories de l'activité (TA) de Vygotsky, plus précisément le modèle de $2^e$ génération de la TA d'Engeström (2000). Les résultats révèlent le rôle des leaders à travers la tension coopération versus collaboration. Dans les équipes émerge un leader qui organise la coopération, mais ralentit l'appropriation des fonctionnalités collaboratives du nouveau système d'information.

**Mots clés :**

Groupe ; leadership ; appropriation ; système d'information collaboratif ; théorie de l'activité# Do leaders facilitate the appropriation of collaborative information systems in teams?

**Abstract :**

This research, focused on small work groups, proposes to examine the role of the leader in the appropriation of collaborative information systems. It is based on a three-year action research conducted during the implementation of the Microsoft Teams collaborative platform at a university. The aim is to understand how groups of students organize themselves to carry out complex projects over a long period of time and with varying distance constraints. Differences in the appropriation of the new tool appeared depending on whether the mode of work was cooperative or collaborative. To understand these different dynamics, we mobilize Vygotsky's activity theories (TA), more precisely Engeström's 2nd generation TA model (2000). The results reveal the role of leaders through the cooperation versus collaboration tension. In teams, a leader emerges who organizes cooperation but slows down the appropriation of collaborative features of the new information system.

**Keywords :**

Group ; leadership ; appropriation ; collaborative information system ; activity theory

# 1. Introduction

La crise du Covid et la parenthèse extraordinairement distancielle qu'elle a provoquée évoquent une mise à l'épreuve et démontrent qu'une organisation est prête (mature) ou non (immature) à transformer ses modes de travail (Laval & Dudézert, 2021). Dans l'enseignement supérieur, on réinterroge ce constat avec une acuité nouvelle et la question du travail en équipe non hiérarchisé en ligne se pose de manière aigüe avec des problématiques diverses.

Alors que les entreprises s'interrogent sur la nécessité de faire muter leur modèle managérial pour s'adapter aux évolutions de l'hybridation du lieu de travail, comprendre l'apprentissage collectif des systèmes d'information (SI) collaboratifs et son impact sur l'organisation du travail de groupes de petite taille est devenu crucial. Dans des situations de travail en groupe, le déploiement de nouveaux outils informatisés permet de renouveler les modalités de création de valeur au sein des activités (Dominguez-Péry, 2011). L'utilité de la technologie pour une meilleure information des membres du groupe en cas d'événements et de changements importants est déjà documentée ; elle conduit notamment à une amélioration de la prise de décision (Tapscott & Caston, 1994). Ainsi pour assurer la continuité pédagogique, les outils collaboratifs prennent subitement une place centrale où le rôle du leader dans leur appropriation collective est finalement peu interrogé par les recherches actuelles.

C'est dans la réalisation de travaux d'étudiants en mode projet avec des contraintes d'éloignement dues à leur statut d'apprentis que l'observation du processus d'usage d'une innovation managériale a permis d'analyser des phénomènes d'apprentissage (David, 2000). L'étude de cas multiples associée à une démarche de recherche-action s'est révélée tout à fait appropriée à la problématique du maintien de la cohésion d'équipe à distance. Etudier les dynamiques des groupes par une mise en situation d'apprentissage collaboratif dans un environnement hybride avec l'introduction d'un nouveau système d'information collaboratif a permis d'interroger a priori deux phénomènes contradictoires. Le premier concerne la forte utilisation d'outils de communication numérique d'Internet par l'ensemble des étudiants durant leurs travaux pédagogiques. Le deuxième concerne leurs difficultés apparentes pour s'approprier les fonctionnalités collaboratives de la nouvelle plateforme Microsoft Teams pourtant conçue dans cette optique collaborative. L'explication se trouve dans une organisation du travail encouragée par un leader persuasif peu propice à la collaboration.

Cette recherche par une approche mixte des dynamiques de groupe dans et hors d'un environnement numérique permet de détecter l'émergence du leadership et d'apprécier son rôle dans l'activité collective médiatisé. En définissant les contours des relations entre l'organisation d'un groupe et son usage de la technologie, elle fait apparaitre les rôles de chacun des membres, notamment d'éventuels leaders, et leur appropriation du SI. Plus précisément, notre recherche s'est focalisée sur la problématique suivante : « *Les leaders facilitent-ils l'appropriation des systèmes d'information collaboratifs dans les équipes ?* ». Pour répondre à cette question, nous mobilisons la théorie de l'activité (TA) de Vygotsky et le modèle de $2^e$ génération de la TA (Engeström, 2001). La TA appliquée au travail et à l'apprentissage (Engeström & Sannino, 2021) met l'accent sur les activités et les tâches que les acteurs effectuent au cours de leur travail plutôt que sur leur personnalité ou leur cognition individuelle. Les résultats de cette recherche permettent de comprendre l'importance du leadership dans l'appropriation de la nouvelle plateforme collaborative MS Teams. C'est dans les équipes présentant un leadership partagé (Luc, 2004) que l'acceptation, l'appropriation collective et enfin les usages du nouvel outil collaboratif se sont révélés rapides et optimaux. L'enrichissement du modèle d'Engeström avec l'élément « leadership » est discuté, car il permet d'évaluer plus rapidement le niveau d'appropriation des équipes grâce à une meilleure compréhension de leurs dynamiques sociales.

## 2. Revue de littérature

Comprendre l'usage d'une technologie collaborative nécessite de saisir la dynamique de la communauté concernée, mais avant toute chose, il faut s'assurer que le collectif observé forme bien un groupe. La littérature classe les groupes en trois grandes catégories : les groupes centrés sur des relations affectives où le but principal des membres est d'être ensemble, les groupes de laboratoire créés artificiellement pour le temps d'une expérience (Leavitt, 1949 ; Bavelas, 1951 ; Flament, 1965 ; Lipitt et White, 1972) et les groupes de travail centrés sur la réalisation de tâches. Il est donc important de commencer par une revue de littérature issue de la sociologie, de la psychologie voire de l'anthropologie (Macdonald, 2016) pour mieux cerner comment les équipes se forment et s'organisent. C'est au cours des années 1930 et 1940 que des psychologues américains ont montré qu'un groupe de travail présente une dynamique propre, au-delà des particularités de ses membres. Ce champ s'intéresse après la Seconde Guerre mondiale à un nouveau champ d'études, la dynamique des groupes. Ce terme désigne l'ensemble des phénomènes, mécanismes et processus psychiques et sociologiques qui émergent et se développent dans les petits groupes sociaux durant leur activité en commun (Lewin 1890-1947). Ces phénomènes ne se manifestent qu'à partir de quatre membres et les groupes humains peuvent être classés en cinq grandes catégories ou les groupes de petite taille fortement structurés sont appelés « Groupes restreints » (Anzieu & Martin, 2019).

Le concept de « popularité » en sociométrie, fait référence au nombre de fois qu'une personne est choisie dans un réseau. L'hypothèse sous-jacente que *« plus une personne a de liens dans un réseau, plus elle est centrale et plus son influence est grande. »* (Mongeau & Saint-Charles, 2005) rends nécessaire une revue de littérature sur le leadership pour comprendre l'organisation des groupes restreints. La prise en compte de la maturité des individus au sein des équipes avant l'émergence d'un leader oriente les lectures vers le leadership situationnel (Hersey & Blanchard, 1974), agile (Hayward, 2021) ou encore facilitateur (Delacroix & Galtier, 2005). Les théories de l'apprentissage collaboratif, développées par des auteurs issus des sciences de l'éducation (Piaget, 1998) ; (Ageyev et al., 2017) ; (Baudrit, 2007) sont des plus éclairantes sur la compréhension des structures collectives d'apprentissage et permettent d'élargir les références afin d'identifier comment les équipes se forment et s'organisent dans un environnement d'enseignement.

Les thèses de la sociologie des outils de gestion peuvent être classées en trois grands types d'approches (Chiapello et al., 2013) « critique » où l'outil est un élément de pouvoir et de contrôle, « institutionnaliste » (Giddens, 2012) ; (Orlikowski, 2010) ; (Carton et al., 2006) ; (F.-X. D. de Vaujany, 2005) s'intéressent plus particulièrement aux dynamiques de transformation sociale qu'accompagnent ou organisent les outils de gestion et « interactionnel » (Weick, 2009), qui met en évidence l'ambivalence de l'outil de gestion conditionné par les jeux d'acteurs (Crozier & Friedberg, 2014). Faisant partie des approches interactionnelles, la théorie de l'activité considère que *« l'outil n'est rien hors du système d'activité »* (Vygotsky, 1978) (Leontiev & Leontiev, 1959), elle permet d'analyser l'appropriation d'un outil de gestion par le jeu des acteurs dans un contexte managérial. Celui-ci peut se présenter sous la forme d'un processus d'apprentissage et de construction de sens rendant l'instrumentation porteuse et génératrice de structure (Grimand, 2006). L'état de l'art commence dans le chapitre 2.1 par la lecture de la structure des groupes grâce à leur réseau de communication. Elle se poursuit dans le chapitre 2.2 par un approfondissement sur la centralité du leader dans le travail coopératif pour se terminer dans le chapitre 2.3 par la prise en compte du rôle croisé des systèmes d'information collaboratifs et du leader dans l'organisation des groupes.

## 2.1 Lire la structure du groupe grâce à son réseau de communication

Le travail en mode coopératif va plus loin que le simple travail en groupe (Derycke, 1991). Il est possible d'en étudier la forme de communication à travers les réseaux locaux :

| Travail en groupe | Travail coopératif | Modèle de communication |
|---|---|---|
| Absence d'interdépendance | Interdépendance positive. | |
| Homogénéité dans la composition du groupe | Hétérogénéité dans la composition du groupe | |
| Un leader par groupe | Partage de la fonction de leadership | |
| Responsabilité de soi-même | Responsabilité de chacun des partenaires | |
| Accent exclusivement sur la tâche | Accent sur la tâche et la gestion de l'interaction pour la réalisation | |
| Méconnaissance des compétences requises pour le travail en équipe | Formation pour le développement des compétences au travail en équipe | |
| Position passive du Manager qui ne prend pas une place d'acteur (médiation) dans l'interaction | Position active du Manager qui observe et intervient dans l'interaction | |

*Tableau 1- Modèle de communication dans une formation collective (Derycke, 1988, p. 9)*

On le voit ici, le passage du travail en groupe au travail coopératif constitue une première étape vers le travail en équipe. Une organisation du travail fait son apparition avec des interactions entre partenaires qui sont gérées par un manager. Les thèmes de formations et compétences font leur apparition, indiquant un lien entre les théories du travail et celles de l'apprentissage. Enfin, le leader qui est ici bien dissocié du manager pourrait ne plus être incarné par un seul acteur, mais par plusieurs. C'est ainsi qu'une organisation plus distribuée de la communication peut permettre au groupe de passer à un mode de travail collaboratif.

Même si une hiérarchie n'est pas préexistante au sein du groupe sous la forme d'un chef d'équipe ou manager, l'émergence d'un ou plusieurs leaders peut figer la forme du réseau de communication en rayon avec le leader en position centrale. Par ailleurs, un certain nombre de caractéristiques différencient le travail collaboratif du travail coopératif. Dans le cadre d'un travail coopératif, la répartition à l'avance des rôles et des responsabilités est absolument nécessaire. Il y aura une répartition claire du travail entre ses participants (Rebetez, 2020). Le leader doit alors tenir compte de la maturité des individus au sein de l'équipe avant d'adopter un style particulier de leadership (Hersey & Blanchard, 1974):

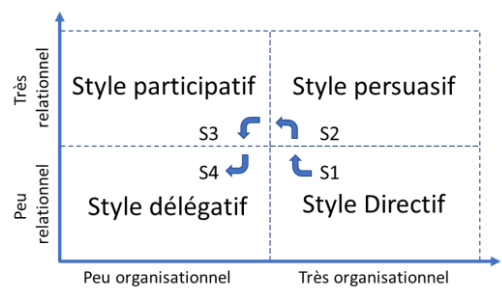

*Figure 1 Les style de leadership en fonction de la maturité des individus (Hersey & Blanchard in Rivai, 2013, p. 161)*

Malgré son style de leadership, le leader est visible dans la forme du réseau de communication et par l'organisation qu'il met en place faisant écho à la structure sociale du groupe.

## 2.2 La centralité du leader dans le travail coopératif

La communication dans le groupe porte l'information sur les modifications de l'environnement *« Autrement dit, elle n'est pas un médium, mais elle est constitutive du processus de changement. Elle assure aussi une fonction dans la jonction entre l'échelle individuelle et collective de ce dernier en favorisant la coordination, en mettant des mots sur les conflits, en explicitant les situations problématiques, etc.»* (Benoit & Méric, 2016, p. 4). Dans le cadre de tâches mobilisant les outils bureautiques d'une manière collective (traitement de texte, tableur, présentation) elle permet de coconstruire un projet en alternant des modes de travail individuel et collectif (De Benedittis et al., 2018, p. 172). Pour étudier ces phénomènes liés aux *« dynamiques des groupes »* (Lewin, 1951), trois grands courants de pensée ressortent dans la littérature. Le courant dynamiste (Balandier, 2018) qui prône une approche expérimentale pour élaborer des théories explicatives où le groupe est défini comme un ensemble de personnes interdépendantes. Le courant interactionniste dont la démarche descriptive nécessite une observation systématique *« armée »* des processus d'interaction entre individus au risque de réduire le collectif à la somme des échanges interindividuels (Bales, 1950). Et le courant psychanalytique dont l'approche clinicienne (Freud, 2015) s'adresse aux psychiatres confrontés aux phénomènes collectifs (Bion & Herbert, 1999) et est aussi mobilisé pour étudier la pédagogie de groupe (Balin, 1992).

Certains postulats avancés à partir des expériences menées sur les groupes restreints dès les années 1950 (Lewin, 1951) mettent en évidence l'importance de la forme du réseau de communication (chaîne, étoile, Y ou cercle) dans la performance des groupes lors de la résolution collective de tâche (Leavitt, 1949). L'idée que la structure du réseau de communication affecte les performances du groupe a été étudiée avec trois formes géométriques différentes d'efficacité plus ou moins grande. Surtout, l'existence d'un lien entre la position centrale dans le réseau de communication et le rôle de leader a été révélée (Bavelas, 1951). Des expériences complémentaires dans des petits groupes ont permis de révéler une forme supplémentaire nommée « réseau tous circuits », qui se différencie des modèles fortement centralisés (Gilchrist et Shaw, 1954). Celle-ci permet de visualiser les dynamiques collaboratives à partir des relations dans un collectif en fonction des liens (niveau de communication ou de relations sociales) entre les nœuds (sujets membres du groupe) :

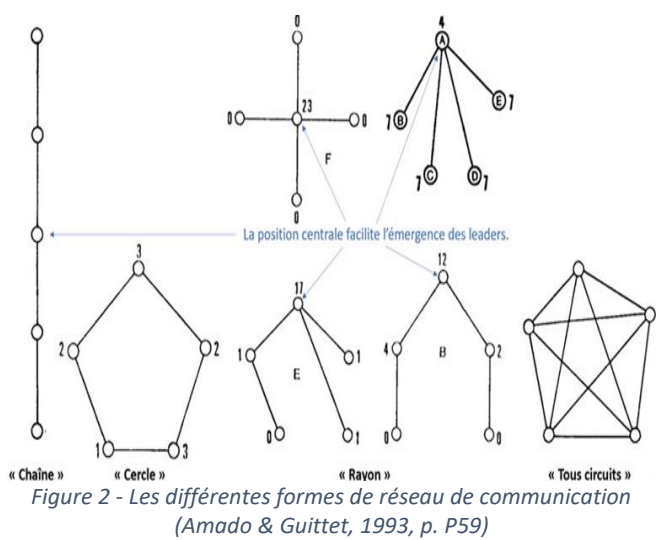

Figure 2 - Les différentes formes de réseau de communication (Amado & Guittet, 1993, p. P59)

La schématisation de la communication ou des relations entre membres d'un même groupe permet la détection des leaders via le concept social de popularité : *« La notion de centralité a été l'une des premières à être utilisée dans l'étude des réseaux. Elle découle du concept de « popularité » qui, en sociométrie, fait référence au nombre de choix que reçoit une personne dans un réseau. L'hypothèse sous-jacente étant que plus une personne a de liens dans un réseau, plus elle est centrale et plus son influence est grande. »* (Mongeau & Saint-Charles, 2005).

On peut alors mesurer l'efficience des groupes en fonction de leur structure sociale et de l'organisation exigée par la tâche (réseau centralisé en étoile et réseau homogène tous canaux) (Flament, 1965).

## 2.3. Les SI collaboratifs sont-ils au service du leadership ?

Même si « *dans le simple appareil de leur technicité, les TIC (technologies de l'information et de la communication) se définissent comme un moyen de relier un ensemble de dispositifs de médiation* » (Benoit & Méric, 2018, p. 2), il ne faut pas s'arrêter à cette simple définition, car les contributions des TIC et plus particulière des systèmes d'information sont aussi à trouver *« dans leur capacité à former des communautés et à en réguler le fonctionnement via le formalisme et la normalisation des plateformes et des autres médias »* (Ibid). Le but principal des usages numériques actuels est de partager à tout moment, de n'importe où, avec qui on veut et comme on veut. La génération qui est actuellement en études supérieures a toujours connu ces possibilités offertes par la technologie grâce aux logiciels mis à disposition gratuitement sur Internet. Sans jugement hâtif sur un quelconque comportement spécifique des jeunes ayant grandi avec le numérique que l'on nomme « digital natifs » (Stenger, 2015) on peut dire que la « génération Y » en France est née et vit *« dans un environnement dominé par des technologies de l'information et de la communication individuelles. »* (Kerneis et al., 2012). Ils ont donc un usage régulier des outils numériques issus du Web 2.0 tels que les réseaux socio numériques[1] (RSN). Ainsi le risque est de voir se développer en parallèle des pratiques de gestion des connaissances spécifiques, un modèle d'une entreprise centrée connaissance, ou organisation 2.0, visant à valoriser globalement ce gène qu'est la connaissance (Dudezert, 2013, p. 66). La réponse à ce danger est l'usage de solutions professionnelles gérées en interne par les entreprises, leur évolution peut être présentée en quatre grands types de systèmes d'information où le Digital Workplace est classé dans les systèmes de collaboration et de partage des savoirs (SGC), car il est centré sur le travail collaboratif, la co-construction, l'échange et la mobilité :

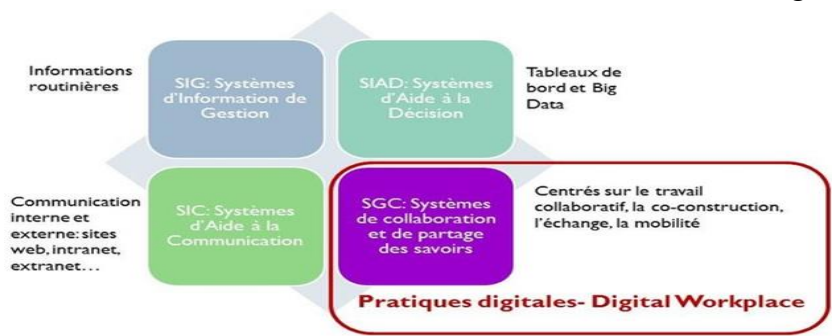

*Figure 3 - La transformation digitale au regard des évolutions des SI (Dudezert, 2018, p. P18)*

Les SGC empruntent un certain nombre de fonctionnalités aux RSN, il est donc intéressant d'étudier si leur appropriation et leur usage en études supérieures et plus largement au travail comportent aussi des similitudes. L'impact social des systèmes d'information collaboratifs doit aussi être analysé au sein même des organisations afin de mieux cerner les contours du management de ces outils de gestion dédiés au travail collaboratif. Les RSN sont présentés comme des solutions pour communiquer au sens large (position centrale) qui par leurs fonctions de partage et de collaboration donnent aux communautés de travail des possibilités d'interaction illimitées via Internet et cela indépendamment de leurs organisations (Shadow IT[2]). Ces comportements dans les entreprises sont très souvent accompagnés par l'utilisation de matériel privé sur le lieu de travail (BYOD[3]). Dans le monde du travail, les outils numériques pourtant utilisés depuis longtemps pour des besoins proches du partage d'informations en mode projet (Groupware) avaient très peu évolué ou du moins beaucoup moins vite. Les systèmes d'information collaboratifs actuels sont issus en grande partie du Web 2.0[4] et des réseaux socio numériques afin de tirer parti de l'appropriation du World Wide Web pour le plus grand nombre. Ils sont censés favoriser la créativité, le partage et l'interaction entre les utilisateurs. Ainsi, ces dernières années des solutions professionnelles sont proposées à des tarifs abordables par les GAFAM[5] en concurrence directe avec d'anciennes solutions de Digital Workplace[6]. Ceux-ci se livrent une guerre sans merci dans le Cloud Computing[7] et s'imposent dans les organisations depuis le confinement du Covid-19 et le développement du télétravail.

---

[1] RSN : *« Un réseau social est un ensemble de relations entre un ensemble d'acteurs. Cet ensemble peut être organisé (une entreprise, par exemple) ou non (comme un réseau d'amis) et ces relations peuvent être de nature fort diverse (pouvoir, échanges de cadeaux, conseil, etc.), spécialisées ou non, symétriques ou non »* (Lemieux, 1999)

[2] Shadow IT : Ensemble d'usages informatiques et d'activités liées aux technologies de l'information qui échappent au contrôle de la direction des systèmes d'information (DSI) de l'entreprise.

[3] Bring Your Own Device : Pratique consistant à autoriser les employés à utiliser dans un contexte professionnel leurs terminaux personnels.

La théorie de la mise en acte : « *L'appropriation est un processus interactif entre les acteurs et des outils, qui engage des prescriptions réciproques* » (Hatchuel, 2018) est une théorie majeure pour l'appropriation des objets. Celle-ci met très tôt en avant le rôle de l'expérience construite par tâtonnements successifs de l'utilisateur, ouvrant ainsi le champ de la sociologie des usages (Perriault, 2008). Dans une approche sociotechnique, la construction sociale de la technologie permet de présenter l'appropriation comme une intervention de l'utilisateur sur les dispositifs (Akrich & Méadel, 2012) où la finalité est de réaliser les livrables attendus collectivement avec ou sans modification du dispositif (utiliser ou non les fonctionnalités existantes de la plateforme numérique collaborative) en modifiant ou non ces usages des outils numériques (moins de réseaux sociaux plus de systèmes de collaboration et de partage des savoirs). Ce rôle de structuration des relations entre acteurs et apprentissage organisationnel suggère que les outils de gestion puissent être des vecteurs d'apprentissage et de changement (Moisdon et al., 1997) ; (David, 1998). D'autres recherches plus récentes s'éloignent d'une vision strictement instrumentale pour en souligner les effets produits sur les dynamiques organisationnelles. (F.-X. D. de Vaujany, 2005); (Grimand, 2006) ; (Oiry, 2011).

La quantité et la rapidité des changements auxquelles les organisations doivent faire face et s'adapter obligent à repenser la dynamique des liens entre les outils de gestion et leurs usagers (Moisdon et al., 1997). Les mouvements qui composent l'action collective organisée ont une matière indissociablement sociale et matérielle. Le tournant matériel permet également de « *repenser l'information et la construction de sens où le corps, les artefacts, l'espace, la matière même de l'information sont au cœur des processus de signification* » (F.-X. de Vaujany & Mitev, 2015). Il y a des relations fortes entre le leadership, la circulation de la communication et la dynamique de travail dans les groupes. Il est judicieux d'étudier l'impact du leadership sur l'appropriation d'un nouveau SI collaboratif lorsque le cadre conceptuel permet de dépasser les modèles classiques du comportement individuel (acceptation des technologies) ou de l'appropriation sociale (sociologie des usages) et pour aller au-delà d'une constatation déjà reconnue de l'effet ambivalent des technologies numériques sur les individus (Weber et al., 2021).

## 3. Cadre conceptuel

La problématique principale de l'étude microsociale du travail groupal médiatisé par les SI est que comme tout outil, ceux-ci restent à l'état d'instrument en devenir tant qu'ils ne sont pas utilisés (l'artefact ne pourra permettre l'apparition d'un instrument qu'à la condition d'être mobilisé dans l'action). Dans le cas précis des systèmes d'information collaboratifs, il faut

---

[4] Les cinq principes qui caractérisent le web 2.0 : 1) Le web en tant que plateforme, 2) exploiter l'intelligence collective, 3) La puissance est dans les données, 4) Le logiciel s'adapte aux différents dispositifs, 5) L'Expérience utilisateur enrichie. (O'Reilly, 2007)

[5] GAFAM : Google, Apple, Facebook, Amazon, Microsoft

[6] Digital Workplace : Atolia, Jalios, Jamespot, Netframe, Talkspirit, Twake, Whaller, Wimi etc.

[7] Le cloud computing (en français, « informatique dans les nuages ») fait référence à l'utilisation de la mémoire et des capacités de calcul des ordinateurs et des serveurs répartis dans le monde entier et liés par un réseau. Les applications et les données ne se trouvent plus sur un ordinateur déterminé mais dans un nuage (cloud) composé de nombreux serveurs distants interconnectés. (*Cloud computing / CNIL*, 2023)

également que les sujets s'emparent de l'artefact dans l'action collective pour la réalisation des tâches pédagogiques ou professionnelles visées pour en faire un instrument commun. L'étude des outils numériques par les enseignants et les étudiants est complexe : *« L'instrument n'est pas un objet neutre, il s'agit d'une entité capable de modifier le comportement de celui qui l'utilise »* (Booms, 2014). Chez Leontiev (1965, 1976, 1984), qui a exploré les possibilités de cette notion d'instrument psychologique, Rabardel (1995) retient l'aspect social et culturel de l'instrument. Celui-ci, chez l'humain, est porteur de modalités d'emploi, issues d'un travail collectif (Rabardel, 1995). La connaissance de l'objet de l'activité (objectif de la tâche) peut aussi être source de complexité dans la collecte des données. Dans le cas précis de l'usage collectif d'instruments dit « numériques », l'activité et parfois même le résultat de l'activité ne sont pas observables directement. Un travail particulier devra être réalisé à partir des traces laissées par l'usage de l'instrument dans l'instrument lui-même (traces informatiques) nécessitant l'accès à ces informations, la connaissance et les compétences pour les analyser.

Enfin le cadre théorique doit obligatoirement prendre en compte la dimension sociale des outils. C'est pour cela que leur rôle de médiation des SI joué en tant qu'artefacts technologiques dans les activités de travail a commencé à être étudié plus complètement. La théorie de l'activité est construite sur le concept de médiation, elle est donc devenue particulièrement fructueuse pour l'exploration à différents niveaux d'analyse des activités individuelles et collectives qui peuvent être vues comme étant fluides les unes dans les autres (Kaptelinin et al., 1995);(Kaptelinin & Nardi, 2006). Pour mieux comprendre l'usage de ces outils de gestion particuliers que sont les systèmes d'information collaboratifs, le choix d'une approche interactionnelle tirée de la sociologie des outils de gestion permet d'étudier non seulement leur état circulant (forme macro) mais aussi leur état inscrit (forme micro). Les travaux de Vygotsky (1997) apparaissent particulièrement adaptés à cette approche. Avec la théorie de l'activité (TA), il considère notamment que l'outil n'est rien hors du système d'activité. Le recensement des travaux les plus récents mobilisant la TA en managements des SI (Gilbert et al., 2013) ;(Ouni, 2002) ;(Quinio, 2009) ;(Weber et al., 2022), en science de l'éducation (Jaillet, 2005);(Fluckiger, 2007);(Booms, 2014);(Gedera & William, 2016) et en sciences de l'information (Abdallah, 2012) ; (Desfriches Doria, 2015) ; (Bourguin & Derycke, 2005) ; (Borgiel, 2017), a permis d'identifier les modèles mobilisables sur notre terrain. Comme le rappelle Abdallah (2012*), « Selon la théorie de l'activité, l'apprentissage de la signification des concepts (objets désactivâtes) se réalise au travers de la compréhension de l'historique et de la dynamique des actions humaines sur ces objets. »* (Abdallah, 2012, p. 3). La TA apporte de nouvelles réponses concernant la problématique de l'appropriation des outils de gestion, ces évolutions ont permis la création du modèle de l'activité collective par Engeström (2001).

### 3.1 La deuxième génération du model de la TA d'Engeström (1987, 2001)

La théorie de l'activité qui est considérée comme prometteuse pour les systèmes d'information et communication *: « La théorie de l'activité propose un cadre conceptuel très largement applicable au champ intrinsèquement pluridisciplinaire des sciences de l'information, notamment en ce qui concerne le développement, l'usage et l'évaluation des systèmes d'information »*. Spasser (1999) est mobilisée entre autres par (K Levan & Vickoff, 2004) pour l'étude des plateaux projets qui peuvent être considérés comme des plateformes numériques collaboratives plus destinées à l'ingénierie technique. Le modèle en forme de triangle dans sa deuxième génération avec les six éléments est complété afin de s'adapter parfaitement à leur contexte de recherche et se révèlera être un outil très bien adapté à l'étude des transformations des sujets ainsi qu'à l'émergence d'activités extérieures et socialisées qui sont accomplies à travers de nouvelles médiations techniques.

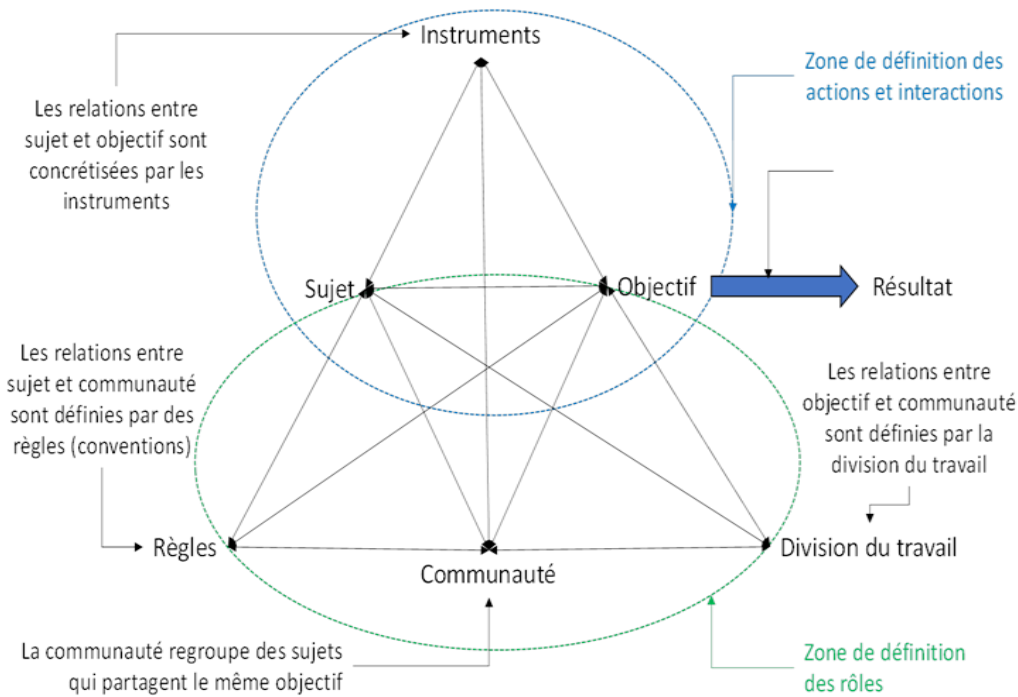

*Figure 4 - Modèle de la TA (Engeström, 2000) adapté par (K Levan & Vickoff, 2004, p. P82)*

La partie haute du triangle représente les actions individuelles telles que définies précédemment, mais la partie basse ajoute des éléments liés au fonctionnement en groupe dans un système d'activité collectif. De plus on peut voir clairement que l'objectif qui est représenté est soumis à des actions multiples *« caractérisées par l'ambiguïté, la surprise, l'interprétation, la création de sens et le potentiel de changement »*. (Engeström, 2001). Elle permet aussi l'analyse de l'action avec des nouvelles médiations et l'étude des développements auxquels ils conduisent. Il peut être avancé que *« transposée au niveau des systèmes d'activités collectives (Engeström, 1999) et des organisations, la théorie de l'activité peut être considérée comme une théorie du changement organisationnel »* (Virkkunen et Kuutti, 2000)

### 3.2 Etudier avec la TA les trois étapes de formation des équipes

L'extension de l'usage d'un outil dans le but d'atteindre un objectif peut avoir de multiples caractéristiques. Dans notre environnement de recherche, nous cherchons à identifier avec tous ces éléments si les sujets forment une équipe, cela nous permet alors de considérer le groupe comme un sujet à part entière pouvant devenir élément d'analyse dans le modèle présenté. Une fois cette hypothèse validée, nous recherchons le rôle de la relation entre le leadership et les outils numériques collaboratifs dans cette transformation. La collecte d'informations sur chacun des éléments, aussi riches soit-elle, ne suffit pas pour répondre à ces questions, une étape supplémentaire est nécessaire. Celle-ci nécessite un agencement particulier de trois éléments du modèle d'Engeström nommées ici triades qui vont faire apparaitre les liens forts susceptibles d'infléchir la définissions des rôles, des actions et des interactions des acteurs qui contribue à l'atteinte du résultat. Voici quelques exemples de triades qui ont été particulièrement utiles pour établir les résultats sur la relation entre le leader, l'organisation de l'équipe et son appropriation du nouveau système d'information collaboratif. Celles-ci sont mobilisées pour classer les résultats dans la quatrième partie en respectant l'ordre d'apparition des étapes successives de formation des équipes.

| | |
|---|---|
| La définition des rôles dans les groupes | |
| | **Triade Communauté-Sujet-Division du travail :** Choix d'organisation des **sujets** (individu) de la **communauté** pour se **répartir le travail** (individu / binôme / coopération / collaboration). Description de la structure des groupes (entretiens, observations et captations). Détection des leaders / « chef de projet » (entretiens, observations et captations) |
| La définition des actions médiatisées dans les groupes de travail | |
| | **Triade Communauté-Sujet-Instrument :** Choix des **instruments** (outils plus ou moins appropriés) par les **sujets** (groupe) pour communiquer avec leur **communauté.** Description de l'équipement numérique et de leurs usages par les étudiants (habitudes). Détection des outils préexistants et évolutions de leur niveau d'usage (entretien, observation, captation). |
| La définition des interactions avec l'instrument prescrit dans les équipes | |
| | **Triade Sujet-Instrument-Objectif :** Les **instruments** (outil collaboratif Ms Teams) permettent-ils aux **Sujets** (équipe) d'atteindre leurs **objectifs** ? Vérifier l'atteinte (notes, entretiens formateurs) des objectifs (livrables documents et oraux). Vérifier le niveau d'usage de la plateforme numérique collaborative (étude des traces). Vérifier l'appropriation de la plateforme numérique collaborative (entretien, observation, captation) |

*Tableau 2 – Trois triades de la TA pour l'analyse du rôle du leadership dans l'appropriation du SI collaboratif*

Avec la théorie de l'activité *« à partir des déterminants de l'activité et en appui sur les acteurs, une réflexion par anticipation devient possible et des préconisations managériales peuvent émaner pour accompagner avec plus de précisions et de subtilités les transformations numériques du travail, réflexion pouvant alors servir de support au moment du déploiement effectif d'une nouvelle technologie numérique. »* (Weber et al., 2021). La théorie de l'activité est donc adaptée à l'élaboration de préconisations managériales dans le cadre d'une recherche-action visant la mise en place d'un nouveau SI.

## 4. Méthodologie

L'observation de la dynamique de communication et de l'utilisation des outils numériques par les groupes est au cœur du dispositif méthodologique. Le processus d'appropriation et d'usage d'un outil numérique et constaté directement avec les acteurs. La démarche de recherche-action a pour objet deux questions intimement liées *« Comment les étudiants travaillent-ils en groupe avec ou sans outils numériques ? Et pourquoi coopèrent-ils plus qu'ils ne collaborent lors des travaux de groupes ? »*. Celle-ci sera engagée dès la deuxième année de recherche afin d'améliorer l'implantation du nouveau système d'information collaboratif MS Teams. Le but est « d'outiller » les étudiants avec des solutions professionnelles pour améliorer leur capacité de travail en groupe aussi bien en présentiel qu'à distance. La dernière année sur le terrain se déroulera pendant les confinements dus au COVID19.

### 4.1 Terrain de recherche et acteurs

L'observation de l'activité de formation dans les locaux de l'école de management s'est concentrée plus particulièrement sur les travaux de groupe. Plus précisément, l'opportunité de participer durant trois ans à l'un des projets transversaux majeurs de deuxième année de master nous a permis d'élargir nos sources d'informations sur les échanges en présentiel et à distance au sein des équipes. Ce projet pédagogique nommé Ciné Corp se déroule tout au long de l'année et consiste en la simulation d'une société de consulting devant répondre à un appel d'offres relatif à une assistance à maitrise d'ouvrage pour une société fictive.

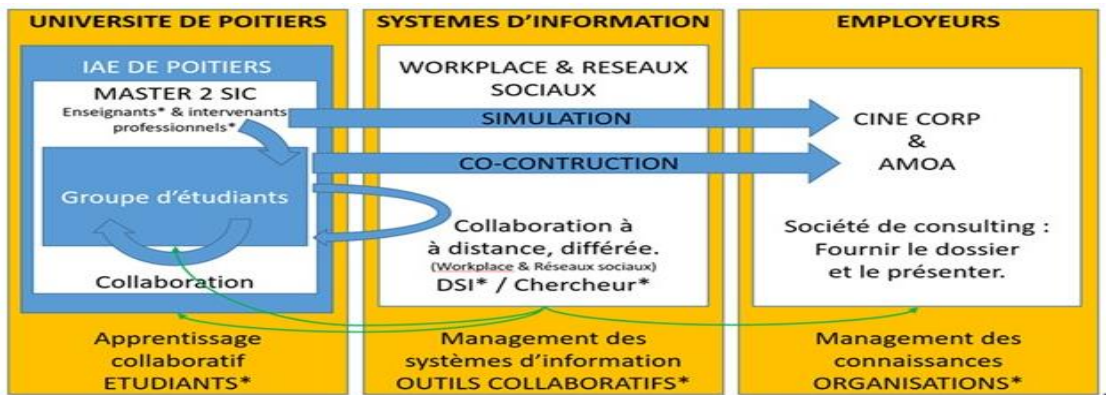
*Figure 5 - Représentation globale des acteurs et outils mobilisés dans le cas Ciné corp*

Les étudiants disposent d'une plateforme collaborative en ligne permettant le travail d'équipe. Administrée par nos soins (accès aux indicateurs d'activité des membres), elle est soumise à un contrat MS Office365 spécifique. Les conversations, fichiers, réunions et applications (formulaires, bloc-notes, etc.) sont rassemblés dans un espace de travail partagé par équipe projet. Une application permet d'utiliser les fonctions de base à partir d'un smartphone :

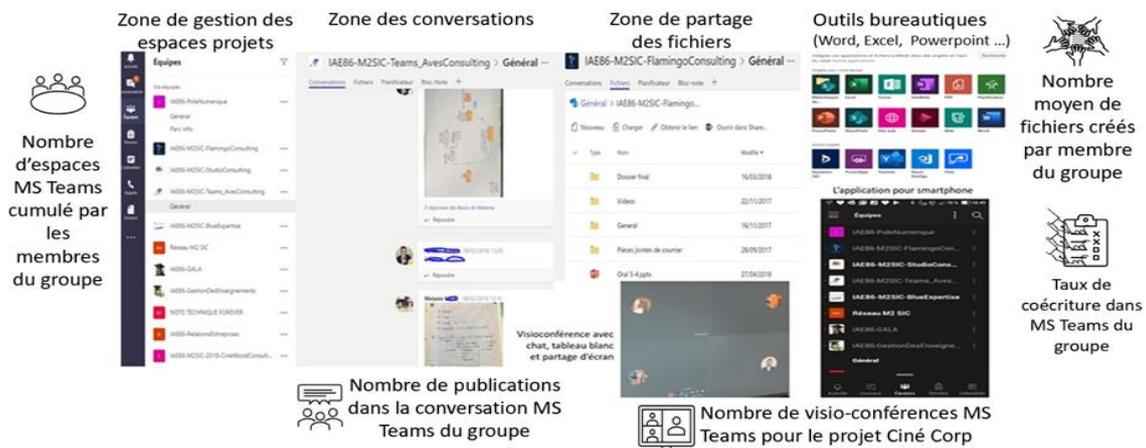
*Figure 6 - La plateforme collaborative MS Teams et les indicateurs d'activité*

Les indicateurs d'activité individuelle dans la plateforme collaborative sont consignés dans des fichiers de log exportables au format .csv à partir de la console d'administration d'Office365. Pour chaque compte informatique il est possible d'extraire les informations suivantes :

- Le compte Office365 online de l'IAE et les activités dans les applications bureautiques
- Les actions générales de gestion de l'espace MS Teams (création, ajout de membres, etc.)
- Les planifications des tâches, notes et autres applications (Planner, OneNote partagé, etc.)
- Les publications (conversation textuel, emoji, photo, document, etc.)
- Les visioconférences (création, le nom des participants, la durée, etc.)
- Les fichiers (création, modification avec version et noms des contributeurs, etc.)

Les quatre fichiers ayant le plus grand nombre de versions avec le plus grand nombre de membres des espaces étudiés seront aussi plus particulièrement étudiés pour voir le niveau de corédaction (Tc) synchrone et asynchrone calculé de la manière suivante :

$$\textbf{Taux de corédaction (Tc)} = \frac{\sum_{n=1}^{4} Nbr\ versions\ doc_n \times Nbr\ rédacteurs\ doc_n}{4 \times Nbr\ membres\ du\ groupe}$$

[8]

---

[8] Les doc n (de n=1 à 4) sont les quatre documents ayant le plus grand nombre de versions, ces fichiers qui ont le plus haut niveau de corédaction sont en général : le dossier final complet écrit en .docx, les présentations orales en .pptx et un fichier de données (budget) en .xlsx. Ceux-ci se révèlent bien souvent les documents fournis à la fin du projet aux formateurs pour l'évaluation finale.

## 4.2 Recueil des données quantitatives et qualitatives

Les techniques utilisées pour collecter les données de terrain sont multiples ce qui permet une collecte de données quantitative et qualitative en prévision d'une triangulation des informations afin de fiabiliser les résultats. Des observations numériques via les relevés de « traces » dans les plateformes collaboratives institutionnelles fournissent les données primaires. Des séries d'entretiens semi-directifs avec les étudiants et les formateurs permettent d'engager l'étude qualitative. L'amélioration globale de l'environnement de travail avec la création de salles de coworking a permis de lancer des expérimentations filmées en autonomie par les étudiants dans nos locaux permettant une analyse très poussée des usages, cette technique est plus fortement mobilisée la dernière année. Les données numériques proviennent des questionnaires nominatifs ainsi que des traces qui ont été collectées dans la plateforme avant d'être traitées. Les entretiens et observations ont été retranscrits intégralement et codés avec NVivo[9].

| Type | Volume de données pour 3 ans | Traitements |
|---|---|---|
| Espaces numériques | 111 créés, accès donné à 90 et 44 espaces MS Teams avec activité collective forte | Repérages et mesure de l'activité Analyse fine avec accès complets aux espaces numériques collaboratifs |
| Quantitative individuel | 33 réponses d'étudiants pour les 9 questionnaires (60 questions) | Analyse quantitative sous forme de matrice et graphique. |
| Qualitative audio | 44 entretiens semi-directifs individuels et 4 focus groups | Retranscription intégrale des 27 heures d'enregistrement et codage dans Nvivo. |
| Qualitative vidéo | Captations présentations orales et auto-captations séances de travail | Visualisation et analyse de 30 heures de vidéos. Retranscription des séquences. |
| Observations | Environ 30 jours dont 25 accompagné d'un formateur | Notes réalisées sur place puis consignées dans le cahier de recherche. |

*Tableau 3 - Récapitulatif du recueil de données (terrains 1, 2 et 3)*

La collecte des données quantitatives a été effectuée à trois niveaux différents. Le premier individuel (niveau d'équipement et usage numérique des étudiants). Le deuxième groupal en rassemblant les informations par équipe (l'activité des différents groupes dans MS Teams, l'organisation dans les phases de travail et de restitution des livrables) (annexe 1). Enfin le troisième au niveau global pour l'ensemble de la promotion (les outils numériques les plus utilisés dans le cadre de leurs études et de leur apprentissage) (Annexe 2).

| PROMOTIONS | EQUIPES | | | PUBLICATIONS | | | VISIOCONFERENCES | | DOCUMENTS | | | |
|---|---|---|---|---|---|---|---|---|---|---|---|---|
| | Nbr espaces MS Teams créés | Nbr moyen membre | Nbr d'accord d'accès | Nbr moyen participant publication | Nombre de publications | Nombre de messages | Nombre de visio | Tps Max Visio | Nombre de documents déposé | Taux cumulé de corédaction | Taux moyen de corédaction | Taux de production de fichiers |
| 2017 M2SIC : | 22 | 3,5 | 22 | 1,4 | 126 | 101 | 7 | 0:15 | 245 | 427,4 | 19,4 | 68,5 |
| 2018 M2CGAO : | 28 | 3,8 | 28 | 0,8 | 60 | 51 | 2 | 0:00 | 229 | 926,4 | 33,1 | 77,8 |
| 2019 M2CGAO* : | 50 | 3,4 | 41 | 2,3 | 549 | 436 | 102 | 15:41 | 607 | 2324,7 | 46,5 | 180,2 |
| *Créé avant confinement : | 37 | 3,3 | | 2,0 | 456 | 360 | 0 | 0:00 | 523 | 1755,1 | 47,4 | 155,0 |
| *Créé pendant confinement : | 13 | 5,8 | | 3,6 | 93 | 76 | 102 | 15:41 | 84 | 618,7 | 47,6 | 25,8 |
| Pour les 3 années : | 100 | 3,5 | 91 | 1,5 | 735 | 588 | 111 | 15:41 | 1081 | 1226,2 | 33,0 | 108,8 |

*Tableau 4 - Relevé d'activité de l'ensemble des projets MS Teams pour sélection*

Les données qualitatives ont nécessité un travail d'analyse des ressemblances / différences afin de révéler les relations entre les différentes variables ayant un rôle dans la dynamique de collaboration des groupes. Le croisement des profils des étudiants par groupe associé à une analyse des entretiens avec le modèle de la TA d'Engeström a permis de dresser des portraits synthétiques des groupes. Les informations sur le leadership sont réparties dans l'ensemble des éléments de codage nécessitant une collecte transversale et globale pour l'ensemble des équipes afin d'en tirer des enseignements généraux. Une attention particulière est portée sur la trajectoire d'appropriation collective du SI dans les groupes en lien avec le type de leadership présent. Ce travail hybride permet de mettre en relation les discours, les observations et les indicateurs d'activités de la plateforme. Les concordances et les contradictions sont relevées grâce aux croisements des données améliorant la fiabilité des résultats.

---

[9] NVivo vers.12 : Logiciel d'analyse qualitative de données pour organiser, visualiser et analyser les données non structurées.

### 4.3 Codage AT et analyse des données dans NVivo

L'ensemble des données collectées concernant le projet transversal sont codées dans NVivo avec les éléments du modèle de la théorie de l'activité. Pour chacun de ces éléments, la création d'un nœud de codage Nvivo va permettre de rattacher les données du terrain correspondantes. Une deuxième phase permettra de regrouper dans un même nœud les parties des éléments constituant une triade significative d'une caractéristique ou d'un comportement particulièrement marquant de l'équipe concernée.

| Nom | Fichiers | Références | Créé le |
|---|---|---|---|
| AT | 25 | 1258 | 02/07/2021 09:13 |
| AT-Communauté | 16 | 267 | 02/07/2021 09:18 |
| AT-Division du travail | 15 | 158 | 02/07/2021 09:18 |
| AT-Instrument | 19 | 452 | 02/07/2021 09:15 |
| AT-Objectif | 19 | 131 | 02/07/2021 09:16 |
| AT-Règles | 8 | 21 | 02/07/2021 09:17 |
| AT-Sujet | 21 | 182 | 02/07/2021 09:16 |

*Figure 7 - Codage des données terrain avec NVivo et le Modèle AT*

Enfin le portrait de l'équipe sera constitué en additionnant les triades, sa forme sera proche d'un récit du fait que nous nous permettons comme l'écrit Paul Ricœur (1984) une narration qui *« recompose les faits de manière créative pour rendre l'action intelligible. Cette reproduction mimétique n'est donc pas une copie passive, mais un réagencement des faits et du temps. »* (Dubied, 2000) se prêtant ainsi à une *« activité mimétique ayant pour objet la mise en intrigue de l'expérience vive »* (Ibid.). L'analyse réalisée dans le logiciel d'analyse qualitative bien que complète semble perdre en partie la composante temporelle si importante pour l'analyse des différentes trajectoires d'appropriations de l'outil. Cet élément crucial sera donc réintroduit dans les portraits des équipes avec le choix d'une forme proche du récit. Un nouveau schéma de l'évolution des groupes - qui deviennent ou pas des équipes - se dessine au fil de l'analyse. Trois triades du modèle AT permettant de suivre précisément l'évolution de l'appropriation de la plateforme numérique collaborative par les groupes d'apprentis sur notre terrain de recherche sont identifiées. Cela permet en les agençant chronologiquement de détecter l'influence majeure du phénomène du leadership sur la vitesse de transformation d'une simple somme d'individualités en un groupe de travail organisé utilisant collectivement le SI.

## 5. Résultats

Les étudiants ont souvent préféré ne pas prendre de risque en s'assurant le confort d'un travail réalisé comme ils en ont l'habitude avec des outils déjà appropriés collectivement et une organisation centrée sur un leader. Cependant, une évolution d'année en année de l'usage de la plateforme MS TEAMS a été constatée avec une forte accélération l'année du premier confinement (annexe n°1). Les indicateurs d'activités collaboratives dans la plateforme numérique ont tous augmenté (annexe n°2) et il a été remarqué par les formateurs du projet Ciné Corp une modification dans l'organisation des groupes avec un phénomène de partage du leadership dans l'équipe ayant l'appropriation la plus aboutie du SI collaboratif. Cette équipe (FOX) aura la capacité de maintenir une dynamique collaborative intense à distance, pour l'ensemble des groupes trois modes d'organisation du travail (coordonné, coopératif et collaboratif) ont été observés. Dans ces modes sera recherchée dans un premier temps l'identité des leaders (5.1) de l'équipe. Dans un second temps, la structure du groupe et la distribution des rôles pour réaliser les tâches seront étudiées (5.2) avec une attention particulière sur l'impact du leader dans l'organisation des actions médiatisées (5.3). Enfin dans un dernier temps le rôle du leader dans l'appropriation de la nouvelle plateforme sera interrogé (5.4).

| TERRAIN 1 : M2 SIC 2017-2018 | | TERRAIN 2 : M2 CGAO 2018-2019 | | TERRAIN 3 : M2 CGAO 2019-2020 | |
|---|---|---|---|---|---|
| **Equipe Ciné Corp n° 1 : STUDIO** | **Equipe Ciné Corp n° 2 : FLAMINGO** | **Equipe Ciné Corp n° 12 : ADAJ** | **Equipe Ciné Corp n° 13 : MAMG** | **Equipe Ciné Corp n° 25 : DAFO** | **Equipe Ciné Corp n° 26 : JALA** |
| Equipe de 5 Structure 2+2+1 | Equipe de 3 Structure 2+1 | Equipe de 4 Structure 2+2 | Equipe de 4 Structure binôme tournant | Equipe de 4 Structure 2+2 | Equipe de 4 puis de 5 Structure 2+2+1 |
| Leadership binôme directif Un membre isolé | Pas de leader déclaré Un membre isolé | Leader délégatif + leader associé directif | Une leader directive | Un leader participatif | Une leader directive Le nouveau spécialisé |
| FBm+Gd+Teams(Tc36) Pas d'appropriation | FBm+Gd+Teams(Tc23) Pas d'appropriation | FBm+Gd+Teams(Tc115) Appropriation partielle | FBm+Teams(Tc133) Appropriation réussie | FBm+MS Teams (Tc63) Appropriation partielle | FBm+MS Teams (Tc122) Appropriation réussie |
| **Equipe Ciné Corp n°3 AVES** | **Equipe Ciné Corp n° 4 : BLUE** | **Equipe Ciné Corp n° 14 : CINEWOOD** | **Equipe Ciné Corp n° 15 : NEOBUSINESS** | **Equipe Ciné Corp n° 27 : FCA** | **Equipe Ciné Corp n° 28 : FOX** |
| Equipe de 4 Structure binôme tournant | Equipe de 5 Structure 3+2 | Equipe de 3 Structure 2+1 | Equipe de 4 Structure 2+2 | Equipe de 4 Structure 2+2 | Equipe de 5 Structure 2+2+1 |
| Leader persuasif | Leader délégative remplacé par un Leader persuasif | Pas de leader déclaré Un membre isolé | Pas de leader déclaré, une leader persuasive cachée | Leader directive + un leader participatif | Plusieurs leaders Leadership tournant |
| FBm+Teams(Tc39) Appropriation partielle | FBm+Teams(Tc57) Appropriation partielle | FBm+Teams(Tc92) Appropriation partielle | FBm+Gd+Teams(Tc49) Appropriation partielle | MS Teams (Tc180) Appropriation réussie | MS Teams (Tc258) Appropriation réussie |

| **Equipe Ciné Corp n° 16 : PHOENIX** |
|---|
| Equipe de 4 ; Structure 2+2 |
| Pas de leader déclaré |
| FBm+Teams(Tc137) Appropriation réussie |

Tc : Taux de corédaction dans MS Teams
Teams : Microsoft Teams
FBm : Facebook messenger
Gd : Google drive

*Figure 8 - Synthèse des résultats issus des treize équipes Ciné corp*

### 5.1 L'identification du leader dans le groupe.

Le rôle du leader s'est révélé prépondérant dans le passage d'une organisation peu structurée ou les étudiants se coordonnent (groupe de trois surtout) à une organisation hiérarchisée avec un leader jouant le rôle de manager. Le leader du groupe est conscient de ce rôle et est bien identifié par les autres membres, ceux-ci n'hésitent pas à le nommer lors des entretiens :

Notes terrain 1 - Ciné Corp 2017-2018, équipe AVES avec un leader persuasif Adam :
Adam : *« Oui il y a un leader moi, c'est moi qui passe mon temps à relancer les autres. »* - Maelle : *« Oui il y a un leader Adam, de toute façon il le sait y'a pas de souci avec ça. »*

Notes terrain 2 - Ciné Corp 2018-2019, équipe NeoBusiness avec un leader persuasif Kevin :
Kevin : *« Oui il y a un leader je pense que c'est moi. […]. J'essaye de faire adhérer un peu l'ensemble. »* - Clara : *« On n'a pas désigné de leader en tant que tel, mais après celui qui prend le plus de décisions et qui a souvent le dernier mot c'est Kevin. »*

Notes terrain 3 - Ciné Corp 2019-2020, équipe FCA avec un leader participatif Jamie :
Jamie : *« Moi je pense que j'ai un peu cette capacité à insuffler des idées on va dire. Ça m'est arrivé dans d'autres projets par contre de dire aux gens : « Là maintenant on va faire ça. »* - Mathéo : *« Après c'est pas quelqu'un qui va prendre le leadership enfin il est pas du tout comme ça Jamie, mais après ses idées sont peut-être plus écoutées que certaines. Moi je le verrais comme ça. Ce n'est pas quelqu'un qui va donner des ordres loin de là. »*

Une équipe très collaborative ne présentera pas de leader formel, MS Teams est utilisé dès le début de l'année pour le partage des documents, les conversations et les visioconférences :

Notes terrain 3 - Ciné Corp 2019-2020, équipe FOX sans leader formel :

Baptiste : « *Je pense que l'avantage du groupe que l'on a, c'est qu'il n'y a pas de leader formel, mais qu'il y a un ensemble de petits leaders qui par moments se présentent.* » - Mélissa : « *Que Teams, Microsoft Office, le téléphone aussi on s'est appelé. Messenger du tout non, on n'a même pas créé de groupe.* » - Jacky : « *La discussion, le partage et la modification de fichiers instantanément et du coup le fait de centraliser l'information, oui de centraliser les fichiers.* »

## 5.2 Le leader valide les rôles dans les groupes.

On constate pour une grande majorité des groupes observés une forte division du travail ainsi que l'émergence rapide d'un leader (figures 8 et 9). Les groupes de quatre sont mieux adaptés aux travaux collaboratifs, la répartition du travail se fait par binômes dont la composition et très souvent organisée par le leader. Les groupes de trois ont été dans l'incapacité de fournir un travail aussi abouti que les autres groupes tant en qualité qu'en créativité malgré des qualités individuelles avérées de leurs membres. Les groupes de cinq ont été confrontés à des phénomènes d'isolement ou de spécialisation de l'un de leurs membres victimes d'une distribution des rôles peu favorable. Le leader encourage les membres du groupe à choisir qui va faire quoi et avec qui. Il organise la division du travail et l'assemblage des résultats. La triade d'analyse est ici composée des éléments **Communauté-Sujet-Division du travail** du modèle de la TA de deuxième génération d'Engeström.

Notes terrain 1 – La première année la promotion n'a pas de formation à la communication et les présentations sont très naturelles ce qui permet d'identifier surement les rôles des membres des groupes et plus particulièrement le leader par sa centralité :

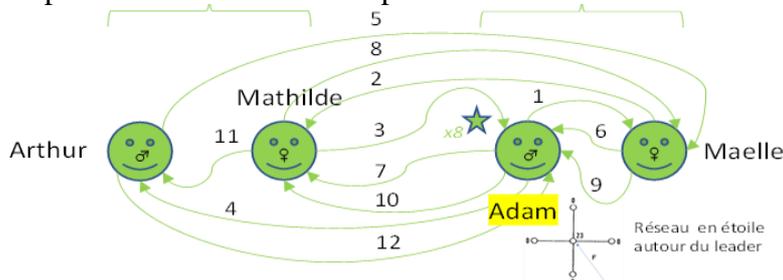

*Figure 9 - Graphique de la structure de la première présentation du groupe AVES.*

Adam a ajouté une règle de fonctionnement à son groupe, a chaque nouvelle partie du projet la composition des binômes doit changer pour s'assurer qu'ils aient tous travaillé ensemble.

Adam : « *J'avais laissé dans les deux autres parties les autres choisir avec qui ils étaient, ils ne savaient pas faire un choix, j'ai donné mon choix et elle a fait en fonction de ça.* »

Notes terrain 2 – La première présentation reste naturelle et source d'information sur le fonctionnement du groupe, car le cours de communication n'a pas encore commencé. Le graphe de l'équipe NEOBUSINESS (figure 9) montre bien les échanges en binômes avec d'un côté Clara et Noé et de l'autre Philip et Kevin. La détection du leader se complique :

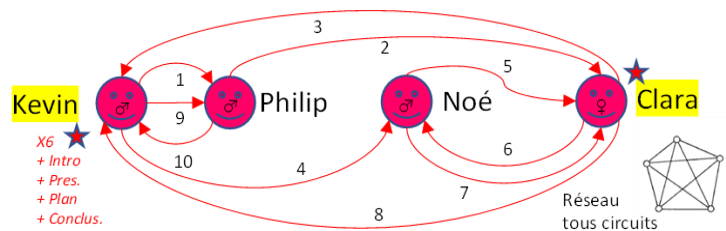

*Figure 10 - Graphique de la structure du groupe NEOBUSINESS lors de la première présentation*

Kevin : *« Quatre c'est bien, on peut diviser le travail en deux groupes pour deux parties. On préfère travailler en binôme c'est plus simple et puis après on centralise tout. »* - Philip : *« Noé et Clara sont plus RH, étude sociale. Et avec Kevin on est plus sur la partie budget en fait. »*

<u>Notes terrain 3</u> – Dès la première présentation les groupes présentent une organisation en deux binômes et deux leaders rendant plus difficile la détection du leader principal. Kevin qui commence la présentation orale (intro) et la finit (conclusion) est clairement désigné comme leader participatif lors des entretiens individuels.

Céleste : *« Ben… on change souvent de binômes, sur certains je suis avec Jamie d'autres avec Fredy, là pour l'analyse de l'audit on change de binômes pour chaque filiale pour justement croiser et ne pas toujours travailler avec la même personne. »*

### 5.3 Le leader organise les actions médiatisées dans les groupes de travail.

La plateforme numérique devant permettre de développer le travail à distance a en fait été très utilisée en présentiel, et a un véritable succès auprès des étudiants lorsqu'elle est associée à un environnement de type coworking. Dans ce cadre de travail, la dynamique collaborative émerge concrètement dans des périodes de brainstorming avant le partage des tâches effectuées souvent en binômes (groupes de quatre) ou parfois individuellement (groupes de trois ou cinq). Dans ces différentes phases de travail, le leader dirige l'action en agissant sur l'environnement et l'ambiance de travail. Ces phénomènes sont validés à la fois par les étudiants lors des entretiens individuels semi-directifs et dans les questionnaires, par les enseignants lors des entretiens individuels et les indicateurs d'activité de la plateforme. La triade d'analyse est ici composée des éléments **Communauté-Sujet-Instrument** du modèle de la TA d'Engeström.

<u>Notes terrain 1</u> - La première année l'IAE ne dispose pas de salles adaptées au travail en petit groupe. Adam dès la première période de travail en autonomie, installe son équipe dans une salle indépendante dans la bibliothèque. Il réagence les meubles afin de travailler face à face deux à deux et utilise le tableau pour établir les stratégies et les montrer au formateur :

Dans ce groupe de quatre, deux filles et deux garçons, tous les étudiants ont un ordinateur portable. Le formateur est inclus dans le groupe et le leader Adam présente les stratégies envisagées sur tableau blanc. Un Word en ligne sous MS Teams est partagé par deux étudiants (Arthur et Mathilde) pour la prise en note des remarques du formateur. Maelle intervient souvent pour poser des questions au formateur.

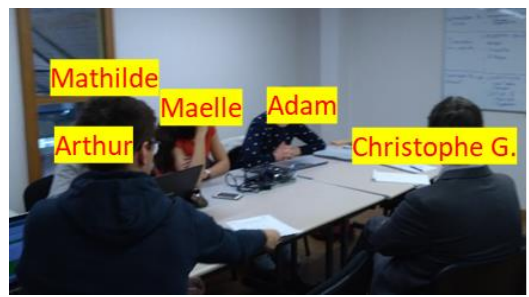

*Figure 11 - Séance de travail du groupe AVES avec intervention du formateur professionnel.*

<u>Notes terrain 2</u> - La deuxième année des salles de coworking vont se mettre en place, mais leur utilisation ne s'est pas encore généralisée. Kevin choisit rapidement de s'installer dans l'une d'elles où il laisse les membres de son équipe échanger librement. Il participe activement aux débats et la décision finale lui revient le plus souvent.

Clara : *« Le choix du nom et du logo ça été assez long 😉 surtout pour le nom de l'entreprise on a tous donné un peu une idée […], mais celui qui en a donné le plus, qui avançait plus ses idées c'est quand même Kevin. D'ailleurs c'était son idée celle-là… »*

<u>Notes terrain 3</u> - La troisième année cinq salles de coworking sont à la disposition des étudiants offrant un espace de travail physique adapté au travail collaboratif. Ce groupe utilise une salle qui comporte deux zones afin de pouvoir s'installer séparément dans chaque partie après une phase de brainstorming ou ils sont tous ensemble. Il est relevé dans la plateforme MS Teams un taux de corédaction de 49 et un taux de production de 8. Ils mobilisent beaucoup les équipements comme l'écran partagé avec ClickShare et MS Teams :

Dans ce groupe de quatre, une fille et trois garçons, tous les étudiants ont un ordinateur portable très récent. Le travail est présenté sur l'écran partagé de la salle de coworking par Jamie et Céleste. Ici la corédaction est effectuée sur un Excel partagé dans MS Teams. La salle a un tableau blanc qui sera utilisé par Jamie lors des brainstormings. La possibilité de travailler séparément en binômes dans cette salle est très appréciée et les documents de travail sont tous partagés dans MS Teams.

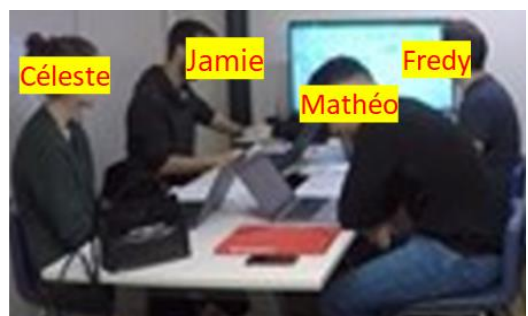

*Figure 12 - Séance de travail du groupe FCA en phase de brainstorming*

Céleste : « *Non je trouve que c'est bien c'est petit, après on a aussi la télé pour regarder des choses, aussi le fait de pouvoir physiquement séparer les binômes je trouve ça bien.* »

### 5.4 Le leader choisit les instruments pour interagir dans les équipes.

Les étudiants ayant déjà développé des stratégies de travail en groupe avec des outils numériques grand public (Facebook Messenger et Google Drive) se partagent très rapidement le travail. Ils ont une démarche plus coopérative que collaborative qui n'est pas propice à l'appropriation d'une plateforme numérique collaborative. Cependant même sans formation, bien qu'ils utilisent peu les fonctionnalités avancées de l'outil numérique collaboratif proposé, ils ont généralisé rapidement son utilisation pour leurs autres travaux en complément de Facebook Messenger. Le leader choisit les outils qui lui semblent d'une part favoriser au maximum les interactions avec son équipe, mais aussi qui lui permettent de centraliser les informations pour se tenir au courant de l'avancée de ses collaborateurs et faciliter le regroupement de leurs productions pour atteindre l'objectif. La triade d'analyse est ici composée des éléments **Sujet-Instrument-Objectif** du modèle de la TA d'Engeström.

Notes terrain 1 - La première année, la plateforme MS Teams de l'IAE connait quelques problèmes, mais Adam s'implique dans sa mise en place, son équipe utilise MS Teams en plus de Facebook qu'ils utilisaient les années précédentes :
Adam : « *Nous je dirais Teams et Facebook, Teams plutôt partage de fichiers et Facebook conversation instantanée.* »
Cela est confirmé par les indicateurs d'activité dans la plateforme MS Teams :
Taux de corédaction = 39 et Taux de production = 9

Notes terrain 2 - La deuxième année, la plateforme est utilisée essentiellement pour le partage de document Word pour le dossier et Powerpoint pour les présentations. Kevin encourage son équipe à utiliser aussi l'application OneNote dans MS Teams pour recueillir et partager les avis et remarques du formateur :
Kevin : « *On utilise OneNote aussi pour prendre des notes rapidement par exemple si Christophe a un retour à nous faire on note dedans et je crois que c'est le premier document que l'on a créé dans Teams.* »

Notes terrain 3 - La troisième année, la plateforme de l'IAE est un outil communautaire, les étudiants considèrent que c'est devenu un réflexe de créer un espace MS Teams pour chaque travail de groupe à réaliser (ici Mathéo précise qu'il utilise sept espaces MS Teams).
Mathéo : « *Actuellement où j'interagis, enfin de créés je dois en avoir sept, je crois un truc comme ça. […] En fait c'est devenu un réflexe, dès qu'il y avait des travaux de groupes* ».

L'équipe FOX s'est approprié collectivement la nouvelle plateforme dès le début du projet :
Mélissa : « *si on n'avait pas eu MS Teams on aurait fait le classique groupe sur Messenger pour partager tout et on aurait perdu énormément de temps à retrouver nos fichiers* »

La troisième année les étudiants terminent leurs projets en confinement, cela permettra de constater une appropriation réussie de la plateforme, car son usage s'est généralisé. Mais il ne faudra pas limiter l'analyse à ce phénomène contextuel : l'éloignement géographique des étudiants sur le territoire Français apparaît tout aussi déterminant :

Mathéo : « *Le confinement n'a pas d'impact sur l'utilisation de Teams dans le cadre du projet Ciné Corp, nous sommes tous en alternance dans des villes différentes, ceci nous a rapidement obligés à utiliser ce nouvel outil pour travailler à distance* »

L'analyse croisée de l'appropriation de l'outil numérique prescrit et du type de leadership dans les équipes a permis d'étudier les évolutions organisationnelles des groupes. Ces résultats encouragent à proposer une évolution du modèle de la théorie de l'activité (Engeström et al., 1999) d'Engeström (1999) en introduisant la notion de leadership afin de mieux appréhender la dynamique collective d'appropriation des systèmes d'information dans les groupes de travail.

## 6. Discussion

Les résultats de cette recherche permettent finalement de discuter sur les rapports étroits entre le comportement d'un groupe primaire ou restreint tel que le définissent Anzieu et Martin (1990) et son usage des technologies numériques collaboratives. Une des difficultés a été de définir les collectifs observés sur le terrain afin de pouvoir sélectionner les cas à étudier. Ainsi, la résistance aux changements définis comme une combinaison à la fois de réactions individuelles, liés à un sentiment de frustration, et collectifs, issus de forces induites par le groupe (Coch & French, 1948) était bien présente sur notre terrain de recherche et a dû être analysée et dépassée pour la réussite du projet d'implantation du nouveau SI.

### 6.1 Le leader fige la structure du groupe en le rendant moins apte aux changements

Même s'il est plus facile de changer les habitudes d'un groupe que celles d'un individu (Lewin, 1951). Nous avons constaté que les groupes où les membres qui se connaissaient et avaient déjà travaillé ensemble opposaient une résistance au changement supérieure du fait d'un attachement plus fort aux normes du groupe. Le leader du groupe joue alors un rôle central pour l'acceptation du changement, l'accompagnement du groupe dans l'appropriation collective des nouveaux outils et l'adoption des nouveaux comportements sociaux.

L'analyse croisée des équipes sur les trois années démontre que la structure des groupes est variable pour la résolution d'une même tâche. Degenne (1966) énonce qu'il existe un apprentissage de la structure qui amène une amélioration des performances et met l'accent sur l'importance du réseau de communication dans la réalisation de la tâche, ces deux éléments sont essentiels dans la compréhension de la dynamique des groupes. Pour Moscovici et Paicheler (1961) la structure cognitive d'un problème favorise l'émergence d'une structure de groupe propre à accomplir de meilleures performances. Alors, comment expliquer que des groupes ayant les mêmes types de tâches à réaliser pour des projets similaires dans un même environnement développent des organisations de travail différentes ? Grâce à l'analyse fine des impacts des multiples formes de réseau de communication détectés lors de la réalisation des travaux, il ressort que le type de leadership peut expliquer en partie les différences d'usage des systèmes d'information collaboratifs.

Dans cette première étape qui consiste à coordonner les individualités, dans le but de les préparer et les outiller pour qu'ils aient la capacité de collaborer, le leader doit avoir l'intuition que la dynamique d'appropriation collective d'un système d'information ne se limite pas à l'utilisation de techniques et d'outils et que son type de leadership doit être moins marqué par le patriarcat, le pouvoir, le commandement et le contrôle et donc être plus propice à l'innovation (Austry et al., 2016). Nous avons constaté au fil des années que beaucoup de chemin avait été parcouru pour que des formes collectives de travail et d'apprentissage jouent un rôle important dans l'enseignement supérieur. C'est aussi l'approche que met en avant Turler (2015) en considérant que le leadership éducatif se retrouve dans l'art de fusionner plusieurs styles et composantes de manière à ce qu'il en résulte une approche collective et cohérente. Cela est totalement cohérent avec notre démarche de recherche-action qui prend en compte qu'apprendre *« c'est étendre l'usage et le but d'un objet en mobilisant des outils pour agir sur lui tout en l'inscrivant dans une communauté de travail »* Engeström (2001). Cette perspective s'inscrit pleinement dans la théorie de l'activité. Cependant elle interroge aussi sur la nécessité de faire évoluer la deuxième génération du modèle de l'activité d'Engeström afin de mieux prendre en compte le phénomène de learship et son rôle dans la dynamique de travail en équipe et l'appropriation collective du système d'information.

### 6.2 Enrichissement du modèle de la 2$^e$ génération de la TA avec l'élément central « leadership »

Le modèle d'Engeström de deuxième génération permet en agençant d'une manière particulière des triades (association de trois éléments) d'analyses d'identifier trois formes d'organisation des groupes de travail différentes et d'identifier l'impact du leadership sur l'ensemble des éléments. Les résultats ont permis de faire ressortir que le leader fédère les sujets pour former la communauté, ajoute des règles pour organiser la division du travail et choisit les instruments pour atteindre l'objectif et obtenir le meilleur résultat. Le modèle original de 2$^e$ génération pourrait donc être enrichi de la façon suivante :

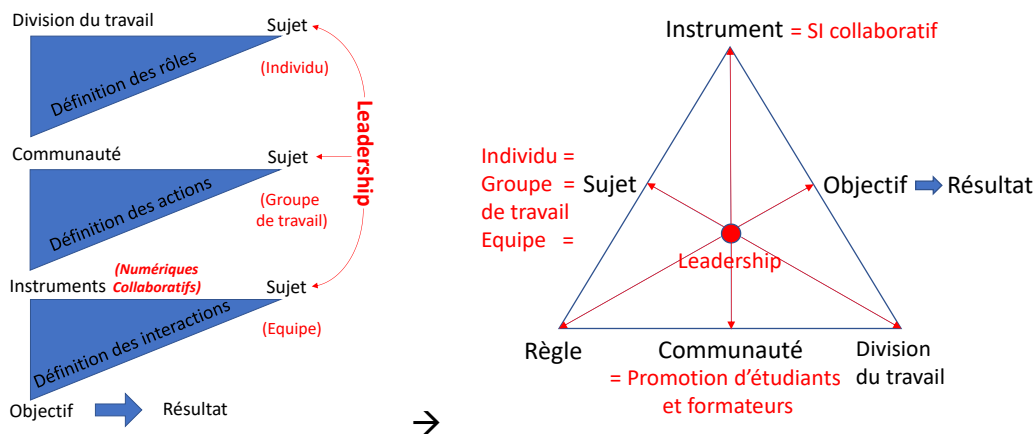

*Figure 13 - Proposition d'évolution du modèle de deuxième génération de la TA d'Engeström*

La première forme de travail collectif nécessite la coordination des individualités (sujet) qui se partagent les tâches (division du travail) en jouant différents rôles pour former un groupe (communauté) et atteindre les objectifs. La deuxième forme de travail nécessite la coopération des membres de la communauté pour constituer un groupe de travail (sujet). Pour cela le leader définit les actions en attribuant les tâches, vérifie les travaux et s'assure de la cohérence des livrables grâce aux outils (instruments). Sous cette forme le groupe de travail devient l'élément d'analyse AT « sujet » du modèle d'Engeström et le SI collaboratif est l'élément d'analyse AT « instrument ». La troisième forme de travail collectif nécessite la collaboration avec des interactions nombreuses entre tous les membres du groupe qui forment alors une équipe (sujet) où le leadership ne semble plus incarné par un individu, mais distribué entre tous les membres. Le système d'information (instrument) en favorisant les échanges directs entre tous les membres, favorise l'atteinte de cette forme de travail. Un leader de style persuasif ou directif peut ralentir l'appropriation du nouvel instrument en figeant le groupe dans une organisation coopérative. La forme collaborative[10] du travail en groupe est très proche de ce que l'on appelle communément des « brainstormings » ou le leader doit adopter un style moins organisationnel et plus relationnel donc délégatif ou participatif. On voit ici l'intérêt d'adopter un style managérial selon la situation de travail tel que précisé par Hersey et Blanchard (1974). Pour atteindre une appropriation complète du nouveau système d'information il faut s'assurer que l'ensemble des équipes ait la capacité de l'utiliser en présentiel et à distance aussi bien pour travailler individuellement, qu'en coopération organisée par un leader que dans une collaboration libre.

## 7. Conclusion

Après une année d'observation, un premier constat fait apparaitre la rareté des périodes d'activité réellement collaboratives dans le projet transversal étudié (Ciné Corp) en lien avec l'organisation des équipes. Un deuxième constat met l'accent sur l'importance du leader et son style de leadership en lien avec les différentes phases de travail collectif en petit groupe (coordonné, coopératif et collaboratif) dans l'appropriation du nouvel outil collaboratif. Les systèmes d'information collaboratifs sont des outils de production et de stockage des données dans les phases de travail individuel, des instruments de partage et de contrôle des connaissances dans les phases de travail coopératif ainsi que des catalyseurs de l'intelligence collective dans un mode de travail collaboratif. Ils permettent aux groupes restreints de maintenir des actions communes spontanées et novatrices ainsi que des relations humaines riches essentielles à la formation et au management des équipes de travail dans des situations d'éloignement spatial et/ou temporel. Ce mode de travail semble être atteint plus rapidement avec le nouveau système d'information collaboratif lorsque le style de management est plus relationnel qu'organisationnel. En effet le style « participatif » qui encourage le partage du leadership entre tous les membres de manière équitable et rotative s'est révélé particulièrement propice à l'appropriation des fonctions collaboratives. Dans les équipes qui ont adopté ce mode de travail ou le leadership et « tournant », l'appropriation du nouveau système d'information collaboratif a été plus rapide et complète. Dans les équipes dirigées par un leader plus organisationnel de style persuasif ou directif, l'usage collectif du nouvel outil est conditionné en grande partie par sa volonté. Le risque est alors qu'il ralentisse son appropriation pour préserver son rôle et l'organisation préexistante de son équipe. Sur le plan théorique, nous proposons d'enrichir le modèle de deuxième génération de la théorie de l'activité d'Engeström (2001) avec le concept de « leadership » comme élément central. Il s'agit de l'adapter aux dynamiques de travail des petits groupes qui recourent à un système d'information collaboratif. Dans la pratique, cela aboutit à la construction d'un outil de détection et de mesure des dynamiques managériales au sein des petits groupes de travail. L'apport managérial se situe également en amont en favorisant l'appropriation de l'environnement numérique collaboratif par l'identification du style de leadership le plus approprié. En aval, une mesure précise de l'impact du nouveau SI sur les équipes et leur nouvelle organisation est également rendue possible.

---

[10] La coopération dans le travail impose une répartition des tâches alors que la force de la collaboration est au contraire d'encourager les participants à travailler tous ensemble à chaque étape.

# Références

# ANNEXES

## 1. Graphique de la trajectoire globale sur trois ans de l'usage de MS Teams

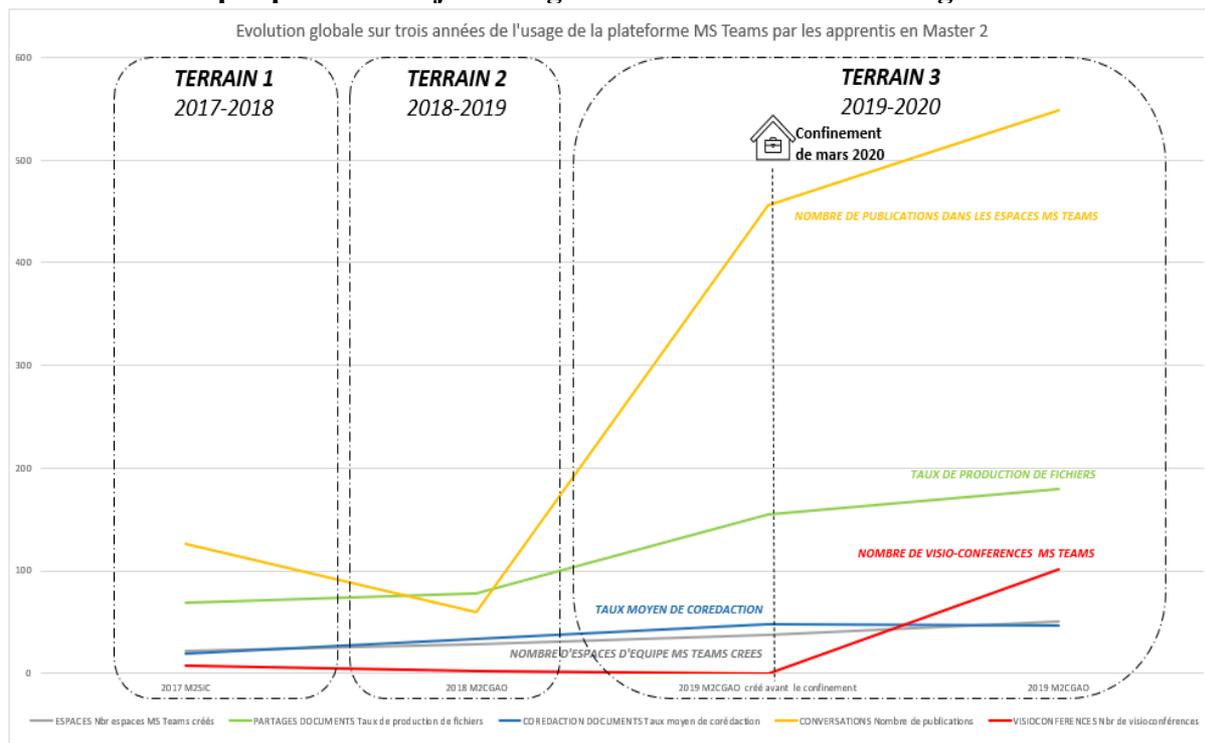

## 2. Graphique de l'activité des équipes Ciné Corp dans la plateforme MS Teams

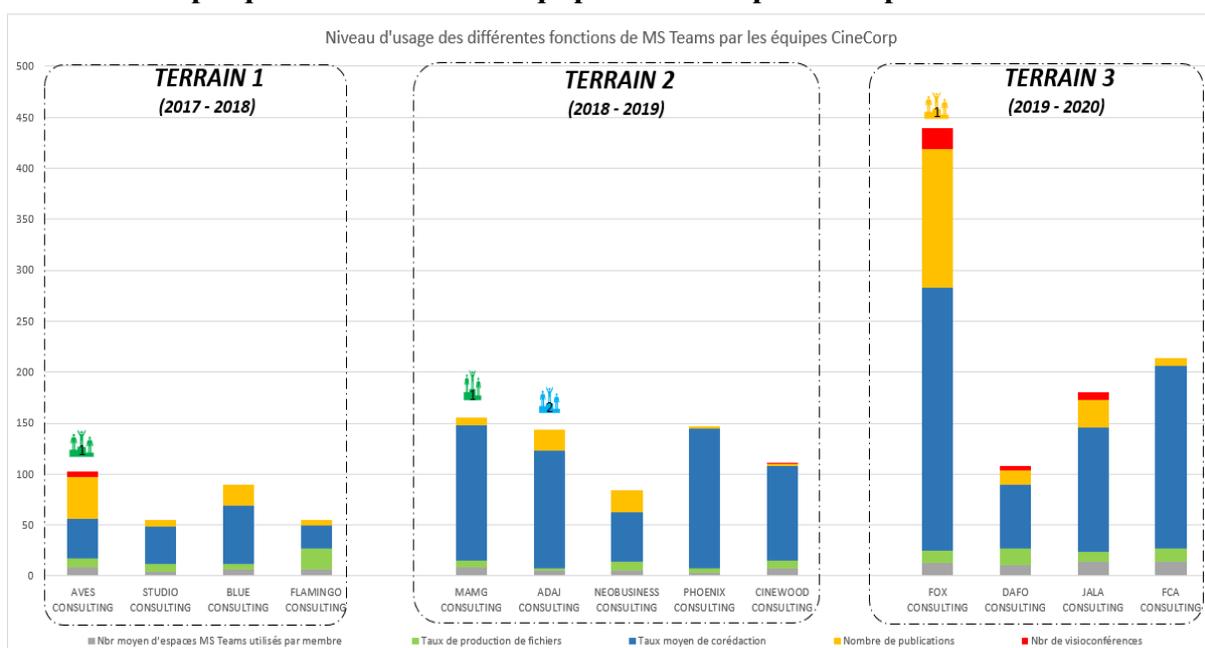